\documentclass[a4paper,fleqn,usenatbib]{mnras}

\usepackage[T1]{fontenc}
\usepackage{ae,aecompl}

\usepackage{graphicx}	% Including figure files
\usepackage{amsmath}	% Advanced maths commands
\usepackage{amssymb}	% Extra maths symbols
\usepackage{epstopdf}
 
\title[S-PASS Cross-Correlation]{Limiting Magnetic Fields in the Cosmic Web with Diffuse Radio Emission}

\author[S. Brown et al.]{
S. Brown,$^{1}$\thanks{E-mail: shea-brown@uiowa.edu}
T. Vernstrom$^{2}$, 
E. Carretti$^{3,4}$, 
K. Dolag$^{5,6}$, 
B.M. Gaensler$^{2, 7}$, \newauthor
L. Staveley-Smith$^{8,7}$,
G. Bernardi$^{9,10}$,
M. Haverkorn$^{11}$,
M. Kesteven$^{4}$,
S. Poppi$^{3}$
\\
$^{1}$Department of Physics \& Astronomy, University of Iowa\\
$^{2}$Dunlap Institute for Astronomy and Astrophysics, University of Toronto, 50 St. George St, Toronto, ON M5S 3H4, Canada\\
$^{3}$ INAF Osservatorio Astronomico di Cagliari, Via della Scienza 5, 09047 Selargius (CA), Italy \\
$^{4}$ CSIRO Astronomy and Space Science, PO Box 76, Epping, NSW 1710, Australia \\
$^{5}$University Observatory Munich, Scheinerstr. 1, 81679 Munich, Germany \\
$^{6}$Max-Planck-Institut fur Astrophysik, Karl-Schwarzschild Strasse 1, 85748 Garching, Germany \\
$^{7}$ ARC Centre of Excellence for All-sky Astrophysics (CAASTRO)\\
$^{8}$ International Centre for Radio Astronomy Research, University of Western Australia, Crawley, WA 6009, Australia \\
$^{9}$Square Kilometre Array South Africa (SKA SA), Park Road, Pinelands 7405, South Africa \\
$^{10}$Department of Physics and Electronics, Rhodes University, P.O. Box 94, Grahamstown, 6140, South Africa \\
$^{11}$Department of Astrophysics/IMAPP, Radboud University Nijmegen, PO Box 9010, 6500 GL Nijmegen, the Netherlands\\
}

\date{Accepted XXX. Received YYY; in original form ZZZ}

\pubyear{2017}

\begin{document}
\label{firstpage}
\pagerange{\pageref{firstpage}--\pageref{lastpage}}
\maketitle

\begin{abstract}
We set limits on the presence of the synchrotron cosmic web through the cross-correlation of the 2.3~GHz S-PASS survey with a model of the local cosmic web derived from constrained magnetohydrodynamic (MHD) simulations. The MHD simulation assumes cosmologically seeded magnetic fields amplified during large-scale structure formation, and a population of relativistic electrons/positrons from proton-proton collisions within the intergalactic medium. We set a model-dependent 3$\sigma$ upper limit on the synchrotron surface brightness of 0.16~mJy~arcmin$^{-2}$ at 2.3~GHz in filaments. Extrapolating from magnetic field maps created from the simulation, we infer an upper limit (density-weighted) magnetic field of 0.03 (0.13)~$\mu$G in filaments at the current epoch, and a limit on the primordial magnetic field (PMF) of B$_{PMF}<$1.0~nG. %This limit on the PMF is comparable to recent limits placed by Planck and the South Pole Telescope, though highly dependent on our assumed model for the population of radiating electrons.
\end{abstract}


%%%%%%%%%%%%%%%%%%%%%%%%%%%%%%%%%%%%%%%%%%%%%%%%%%

%%%%%%%%%%%%%%%%% BODY OF PAPER %%%%%%%%%%%%%%%%%%

\section{INTRODUCTION} 

The evolution of the diffuse intergalactic medium is difficult to observationally constrain, owing to its low density and temperature. The T=10$^5$-10$^7$~K warm-hot intergalactic medium \citep[WHIM][]{dave01,cen06} is extremely faint at most wavelengths, though it contains as much as half the total baryon content of the local Universe \citep{breg07}.  Direct detection of this gas is extremely difficult, though some recent claims have been made \citep{cant14,mart15,ecke15}. Indirect methods have been proposed to try and infer the presence of the WHIM, such as the pressure balance in Giant Radio Galaxy lobes \citep[e.g.,][]{mala13} and Fast Radio Burst dispersion measures \citep{akah16,fuji17}. Another indirect method is to observe the non-thermal radio synchrotron emission generated by the relativistic plasma believed to coexist with the WHIM \citep{kesh04,pfro06,ryu08,skil08,aray12,vazz14,vazz15a,vazz15b}. If this plasma is detected at radio wavelengths, it can constrain the origin of the large-scale magnetic fields that permeate the intergalactic medium of galaxy clusters \citep[see][for a review]{brun14}, differentiating, e.g., between the interaction of radio and star-forming galaxies with the WHIM \citep[e.g.][]{donn09} and a primordial magnetic field \citep{mari15}. Observationally inferring the presence of magnetic (B) fields outside of dense clusters of galaxies is extremely difficult \citep[e.g.][]{xu06}, and B field estimates from simulations can vary from 0.2 $\mu$G \citep{cho09} to 0.005~$\mu$G \citep{donn09}, depending on the models used for seeding and amplifying the fields. Any measurement of these currently unconstrained values would provide a useful constraint for differentiating between the possible models. 

The presence of a faint radio ``Synchrotron Cosmic Web" (which we will call simply cosmic web for short) that traces the WHIM, has yet to be observationally confirmed \citep{wilc04,vazz15b}. The physical mechanisms producing this synchrotron emission are the acceleration of relativistic particles by shocks and turbulence, and recent cosmological simulations have now focused on tracing these mechanisms in the filamentary regions of the cosmic web \citep{aray12,vazz14,vazz15a,vazz15b}, and they predict surface-brightnesses up to a few $\mu$Jy/arcmin$^2$ in the largest filaments at 1.4~GHz, which is currently only attainable through the statistical analysis we propose below.

Besides intrinsically low surface-brightness, diffuse emission from the Milky Way presents yet another obstacle for the detection of the synchrotron cosmic web. Synchrotron emission from the Milky Way dominates the cosmic web at all frequencies and scales, though surface-brightness fluctuations decrease at smaller angular scales \citep{lapo08}. Unlike the foreground issues in cosmic microwave background (CMB) analysis, this foreground has a similar frequency behavior to the cosmic web, and thus cannot be removed with template matching at other wavelengths \citep[e.g.][]{gold11}. At the present time, statistical detection of the cosmic web may be the only option to get around the problem of confusion from both the Galaxy and background extragalactic sources. 

One possible method of statistical detection is cross-correlation. Cross-correlation is a technique to statistically detect emission below the noise level of a map (both instrumental and source noise) . The widest use of this technique has been on the cosmic microwave background (CMB), where it has been used to detect the late Integrated Sachs-Wolfe (ISW) effect \citep[e.g.][]{crit96,boug04,nolt04,mcew07,liu11}. Cross-correlation can be performed in many ways, and has been attempted successfully in real \citep[e.g.][]{nolt04}, harmonic \citep{afsh04}, and wavelet \citep{mcew07} space. For our purposes, cross-correlation can be used to infer the presence and average surface-brightness of the synchrotron cosmic web \citep[e.g.][]{brow10,brow11b,vern17}, provided that we have a plausible model for the spatial distribution of the diffuse radio emission. This method was first attempted on a small region of the northern sky using single dish observations at 1.4~GHz \citep{brow10}, and more recently utilizing deep Murchison Widefield Array (MWA) data at a frequency of 180~MHz \citep{vern17}. Both studies used the distribution of galaxies (binned in redshift) as tracers of large-scale structure.  In the case of \cite{vern17}, they further modified the galaxy distribution (galaxies per pixel) to model the distribution of the diffuse radio emission through convolution at Mpc scales, and utilised the scaling relations derived in \cite{vazz15b} to derive magnetic field upper limits. One limitation to this approach is the lack of a detailed spatial model of how the radio emission correlates with large-scale structure (beyond linearly), something that the current work attempts to overcome using constrained magnetohydrodynamic simulations of the local Universe. 

In this paper we cross-correlate the radio continuum map from The S-Band Polarization All-Sky Survey (S-PASS) with a model of the synchrotron cosmic web obtained through a cosmological simulation of the local Universe. In $\S$2 we describe the S-PASS data, the model of the cosmic web, and our pre-processing of the images. $\S$3 outlines the cross-correlation procedure, results, and how we compute upper-limits on the surface-brightness of the cosmic web, and in $\S$4 we discuss the implications of our limits for theories of cosmic magnetism and particle acceleration in large-scale structure, as well as prospects for direct detection with upcoming Square Kilometre Array pathfinders. 

\section{The Data}  
	 
\subsection{S-PASS Radio Data}	 
\begin{figure*}
\begin{center}  
\includegraphics[width=8cm]{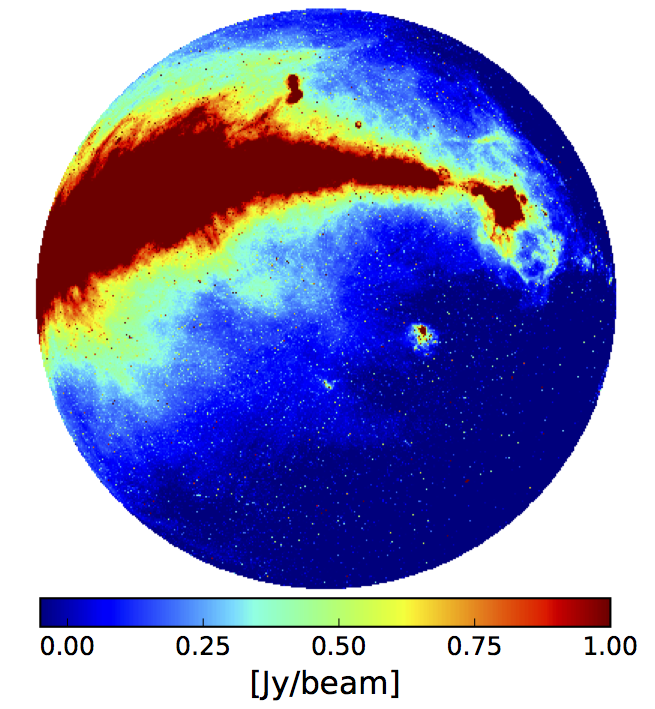}
\includegraphics[width=8cm]{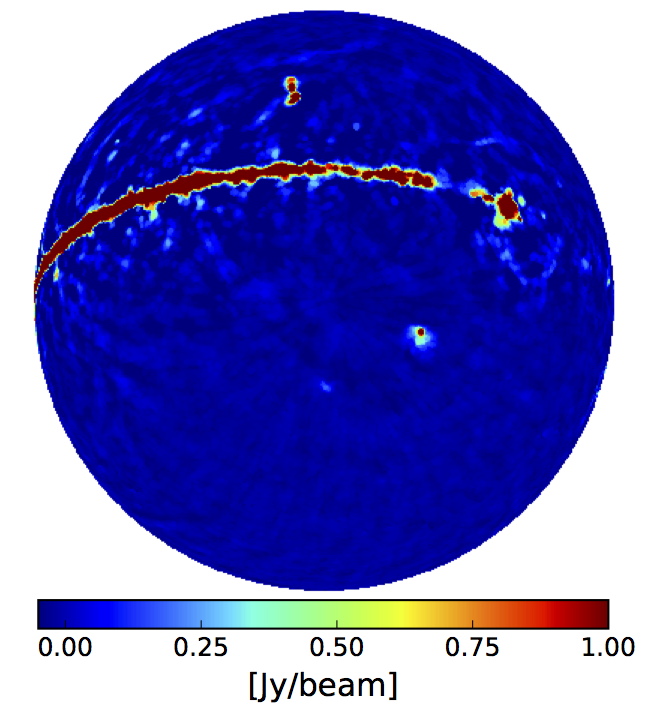}
\begin{footnotesize}
\caption{\label{spass} Left: Original S-PASS image orthographic projection (half the sky) centered on the South Celestial Pole; Right: Residual small-scale emission after subtracting a median-weight filtered image (300$^{\prime}$ circular filter), and applying a spatial scale filter \citep{rudn02} that removes emission less than 27$^{\prime}$ in size.}
\end{footnotesize}
\end{center}   
\end{figure*}

The S-Band Polarization All-Sky Survey (S-PASS) is an 2.3~GHz continuum polarization survey of the southern sky using the Parkes 64~m telescope operated by CSIRO Astronomy \& Space Science in NSW, Australia \citep{carr13a}. The current paper only utilizes the total intensity image produced from the survey. The 2.3 GHz system at Parkes has a resolution of 9$^{\prime}$, however final S-PASS maps have a resolution of 10.75$^{\prime}$, after a convolution with a Gaussian kernel with FWHM=6$^{\prime}$ applied during the map-making process (details will be presented in the forthcoming S-PASS survey paper, Carretti et al., in prep.). With a sensitivity of 1~mJy~beam$^{-1}$, and high sensitivity to large-scale diffuse emission, S-PASS is ideally suited for detecting synchrotron emission within the cosmic web. The point-source confusion limit is 13~mJy/beam in the total intensity images \citep{meye17}. The survey data was converted into Healpix format \citep{gors05} with N$_{side}$=1024, which has an approximate pixel size of 3$^{\prime}$ (Fig. \ref{spass}, Left). We have already utilized the S-band Polarization All-Sky Survey (S-PASS) to detect a synchrotron bridge connecting the cluster centre to the major outgoing merger shock in the massive cluster Abell 3667 \citep{carr13}, which demonstrates the ability of S-PASS to detect faint extragalactic synchrotron emission associated with large-scale structure.

The S-PASS image is dominated by large-scale Galactic emission, so we have subtracted out a 300$^{\prime}$ median-weight filter. The filter size was chosen to optimize the removal of the Galactic foreground without subtracting too much of the potential cosmic web emission, which we estimated from our model below. After this filtering, Galactic and extragalactic point-sources remain, as well as diffuse emission less than $\sim$300$^{\prime}$ in extent. In order to remove the point sources, we further applied a circular spatial-scale filter of diameter 27$^{\prime}$, chosen to be 2.5x the beam to remove bright compact sources with $<1\%$ residuals \citep{rudn02}. Figure \ref{spass} (Left) shows the resulting map, which has an rms noise of 26~mJy/beam (mostly confusion from residual compact sources and Galactic diffuse emission) in the clean unmasked regions (see \S 3.1). It should be noted that diffuse emission of any spatial scale {\it below the noise level of the map} will not get removed by the median or spatial-scale filters \citep{rudn02}. We account for any missing cosmic web flux in our analysis of the upper limits below.  

\subsection{Synchrotron cosmic web Model}
In order to spatially resolve filaments of galaxies, which have a typical width of 1-2 Mpc \citep[e.g.][]{ryu03}, we restrict our redshift range of interest to $z<0.048$ in order to have at least $2\times10^{\prime}$ S-PASS beams across a 1~Mpc filament\footnote{Resolving filaments is necessary to distinguish the diffuse filamentary IGM radio emission from compact radio/star-forming galaxies within the filaments}. We have therefore adopted as a model the simulations of \citep{dola04,dola05} and \cite{donn09}, which trace the amplification of magnetic fields, along with the thermal gas, in a constrained simulation that reproduces local large-scale structure. The simulation used initial conditions based on the IRAS 1.2~Jy galaxy survey \citep{fish95}, and reproduces local clusters and super-structures  with sizes $>$3~Mpc (e.g., Virgo, Coma, Centaurus, Hydra, Perseus, A3627), including some smaller structures such as the merging substructure within the Coma cluster \citep{donn10}. The source of the radiating CR electrons in the simulation are secondary electrons/positrons (CRe$^{\pm}$) created from collisions between CR protons and the thermal protons of the WHIM \citep{denn80,blas99}. The motivations for using the secondary model are twofold; 1) the CR protons should be shock-accelerated during the buildup of large-scale structure (LSS) and live for a Hubble time, making their presence in filaments expected at some level \citep[e.g.][]{blas07}, and 2) the number density of these CRe$^{\pm}$ should be low in filaments, so using this model should produce a conservative upper limit on the magnetic fields.

\begin{figure}
\begin{center}  
\includegraphics[width=\columnwidth]{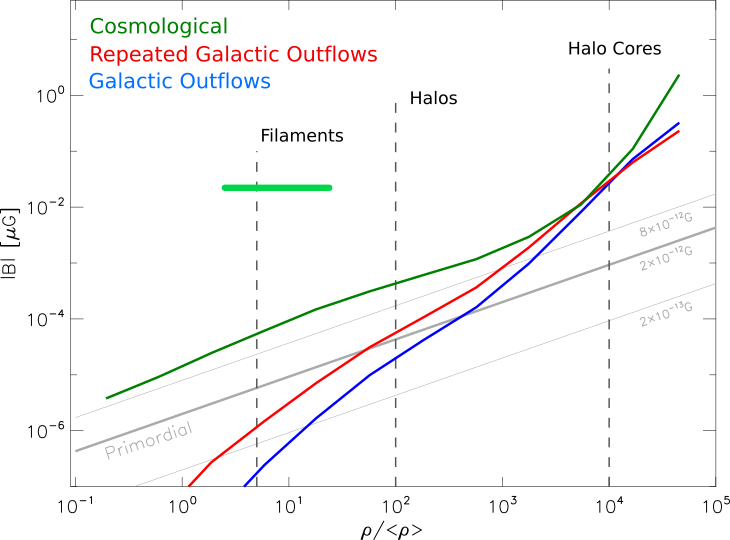}
\begin{footnotesize}
\caption{\label{mods} Plot of average magnitude of magnetic field as a function of fractional density for the models used by Dolag et al. (2004, 2005) and Donnert et al. (2009). This work focuses on the {\it cosmological} model, which predicts the highest values within filaments. The horizontal green bar is the 3$\sigma$ upper limit derived in this work. $<\rho>$ is the average density of the simulation.}
\end{footnotesize}
\end{center}   
\end{figure}

\cite{donn09} used different models for seeding the magnetic fields at high redshift, and traced the amplification of these fields during the collapse of the local structures.  These simulations have been used to model the giant radio halo of the Coma cluster \citep{donn09}, Faraday rotation from the local cosmic web \citep[][ also through cross-correlation]{stas10},  and the origin of fast radio bursts \citep[FRBs,][]{dola15}. Figure \ref{mods} shows magnetic field strength as a function of density for the various magnetic field models used with the simulation. The magnetic field values in filamentary regions for all models in Fig. \ref{mods} are lower than the work of, e.g., \cite{ryu08} and \cite{cho09}, that predict B$\sim$0.001~$\mu$G, though there is no discrepancy in dense cluster cores.

\begin{figure*}
\begin{center}  
\includegraphics[width=8cm]{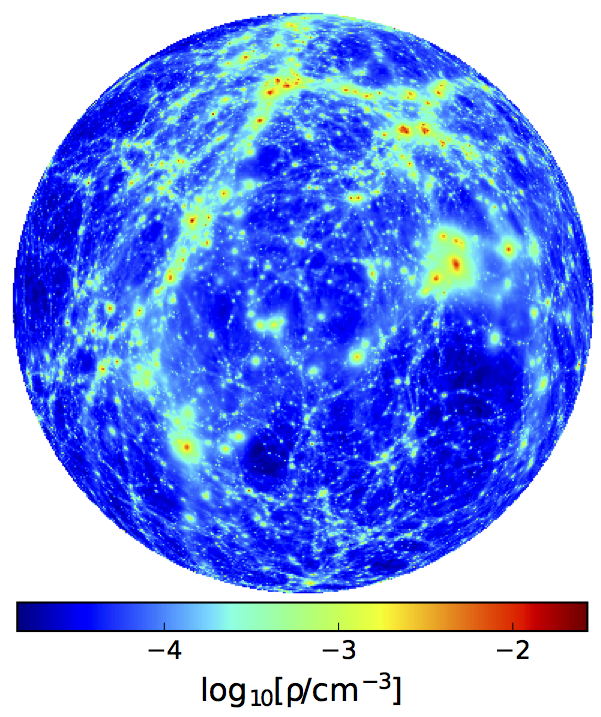}
\includegraphics[width=8cm]{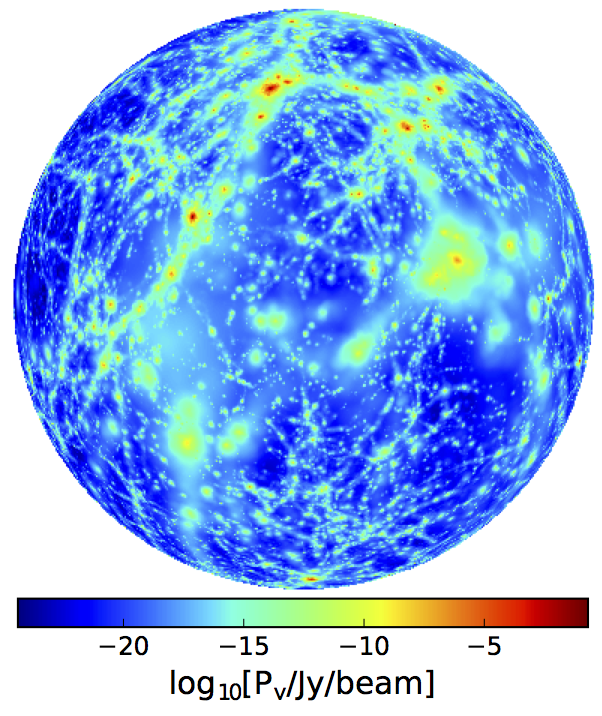}
\begin{footnotesize}
\caption{\label{models} Left: Electron density map ($log_{10}$(n$_e$) [cm$^{-3}$]) from the {\it cosmological} model of \citep{dola04}; Right: Radio synchrotron map ($log_{10}$(I)  [mJy~arcmin$^{-2}]$) at 1.4~GHz from the {\it cosmological} model.}
\end{footnotesize}
\end{center}   
\end{figure*}

The model that we explore in this paper (which we call {\it cosmological}) is the one presented by \cite{dola04}, which uses a uniform cosmological magnetic field of 10$^{-11}$~G seeded at high redshift and evolved to the present epoch (green line in Fig. \ref{mods}).  This primordial magnetic field (PMF) is an order of magnitude lower than the upper limits placed by Planck \citep{plan16}, though in Fig. \ref{mods} we can see that this model predicts the highest values for magnetic fields within filaments, and is the most optimistic model for detection given the limited sensitivity of the S-PASS maps. The other two models in Fig. \ref{mods} consider seeding magnetic fields through dipolar galactic outflows at either a single epoch ({\it Galactic Outflows}) or multiple epochs ({\it Repeated Galactic Outflows}), both predicting significantly weaker magnetic fields in filamentary regions compared to the {\it cosmological} model. The solid grey lines show the results of simple adiabatic compression of a primordial field, for reference. Fig. \ref{models} shows the electron density and 1.4~GHz radio emission from this model. The radio emission in the cluster cores is roughly what is currently observed in the Coma cluster \citep[e.g][]{brow11a}, though the simulation shows clearly more diffuse emission in the filamentary regions; this is the as yet undetected synchrotron cosmic web. 

\section{Cross-Correlation Analysis \& Upper limits}

We perform a cross-correlation of the S-PASS and the cosmic web map by using {\sc anafast} implemented in the {\sc Python} module {\sc Healpy}. We first calculate the pseudo cross-power spectrum \citep{hivo02} through  {\sc anafast}'s spherical harmonic decomposition and then calculate a cross-correlation function from the measured $C_{l}$ values from, 

\begin{equation} CCF\left(\theta\right) = \sum_{l} \frac{2l+1}{4\pi}C_{l} P_{l}\left( cos\theta \right),
\end{equation}

\noindent where the $P_{l}$ are Legendre polynomials. The error bars used below are calculated using cosmic variance only, and are given by $Var(C_{l}) = 2C_{l}^{2}/f_{\mathrm{sky}}(2l+1)$, which should be regarded as lower limits.  Before we perform the cross-correlation, we need to construct the appropriate mask, as well as measure the likelihood of spurious correlations in the data.

\begin{figure}
\begin{center}  
\includegraphics[width=8cm]{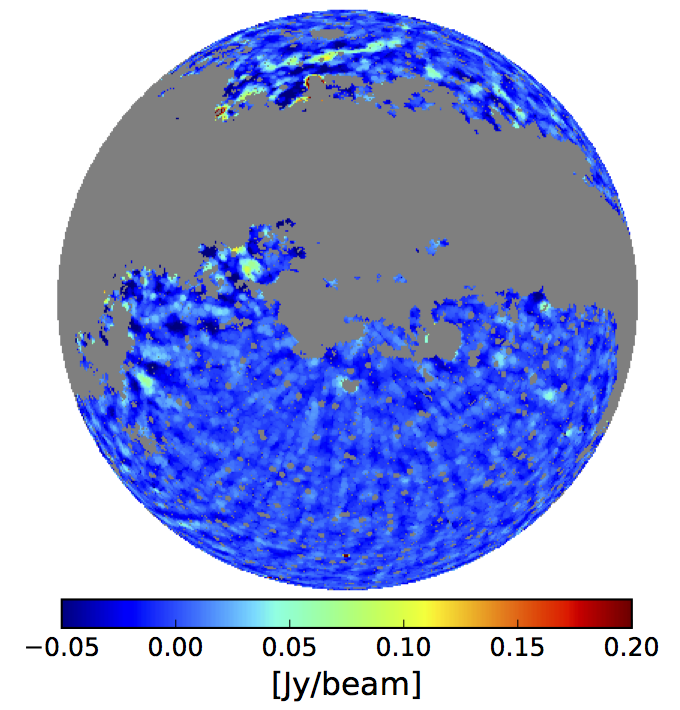}
\includegraphics[width=7.5cm]{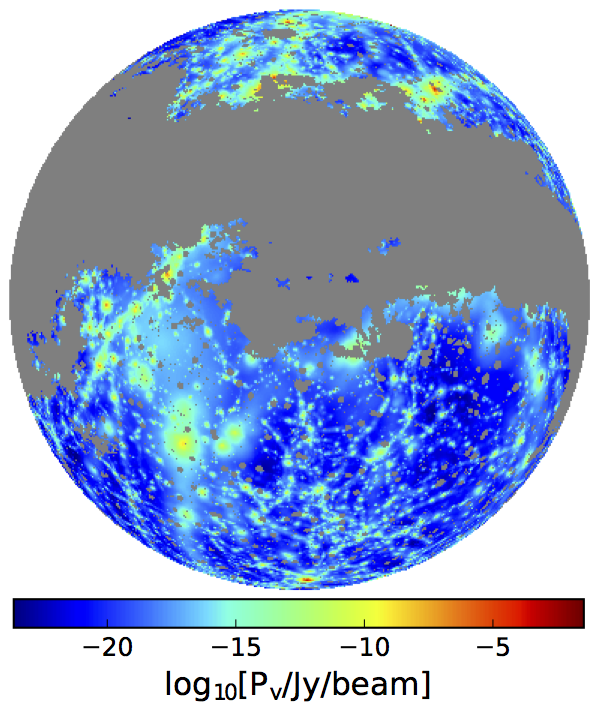}
\begin{footnotesize}
\caption{\label{mask} Top: S-PASS image after all the filtering and with the mask applied. Note that the scale has been changed compared to Fig. \ref{spass} in order to bring out low-level features. Bottom: The radio emission from the simulation with the same mask applied. The total unmasked fraction of the sky is $f_{\mathrm{sky}} = 26\%$.}
\end{footnotesize}
\end{center}   
\end{figure}

\subsection{The Mask} We are only interested in correlations between the two maps that correspond to low-density regions of the cosmic web, and must therefore mask out other possible sources of correlation. Our first consideration is the Galactic plane, which needs to be excluded for several reasons. First, the S-PASS map (Fig. \ref{spass}) shows very large-scale diffuse emission in the plane which is for the most part not extragalactic. In addition, the model of the cosmic web was based on constrained simulations of the local LSS, which used initial conditions based on the IRAS 1.2~Jy galaxy survey \citep{fish95}; these initial conditions could be contaminated by Galactic dust extinction, resulting in an anti-correlation \citep[see][]{dola04}. We therefore applied the {\sc Planck} CMB mask  \citep{plan16} which masks out the Galactic plane and many of the brightest extragalactic point-sources. Residual emission remained due to the Galaxy and the bright extragalactic source Centaurus A  \citep{feai11}, so we further masked regions where the filtered S-PASS map was $>$0.1 Jy/beam. We have also masked out regions with extinction E(B-V)$>$0.1 from the Schlegel et al. (1998) maps, as well as galactic latitudes $|b|<$10 degrees.

As we are interested in only the filamentary regions of the cosmic web, we have also masked out regions where the thermal electron density is greater than 10$^{-3}$~cm$^{-2}$. The results are shown in Fig. \ref{mask}. The total unmasked fraction of the sky is $f_{\mathrm{sky}}$=26\%. %We do not include this fraction when calculating the error-bars on Equation 1, but the reduced fraction of the sky naturally contributes to the scatter in random correlations used below to calculate the upper limit. 

\subsection{Random Correlations} 
\begin{figure*}
\begin{center}  
\includegraphics[width=14cm]{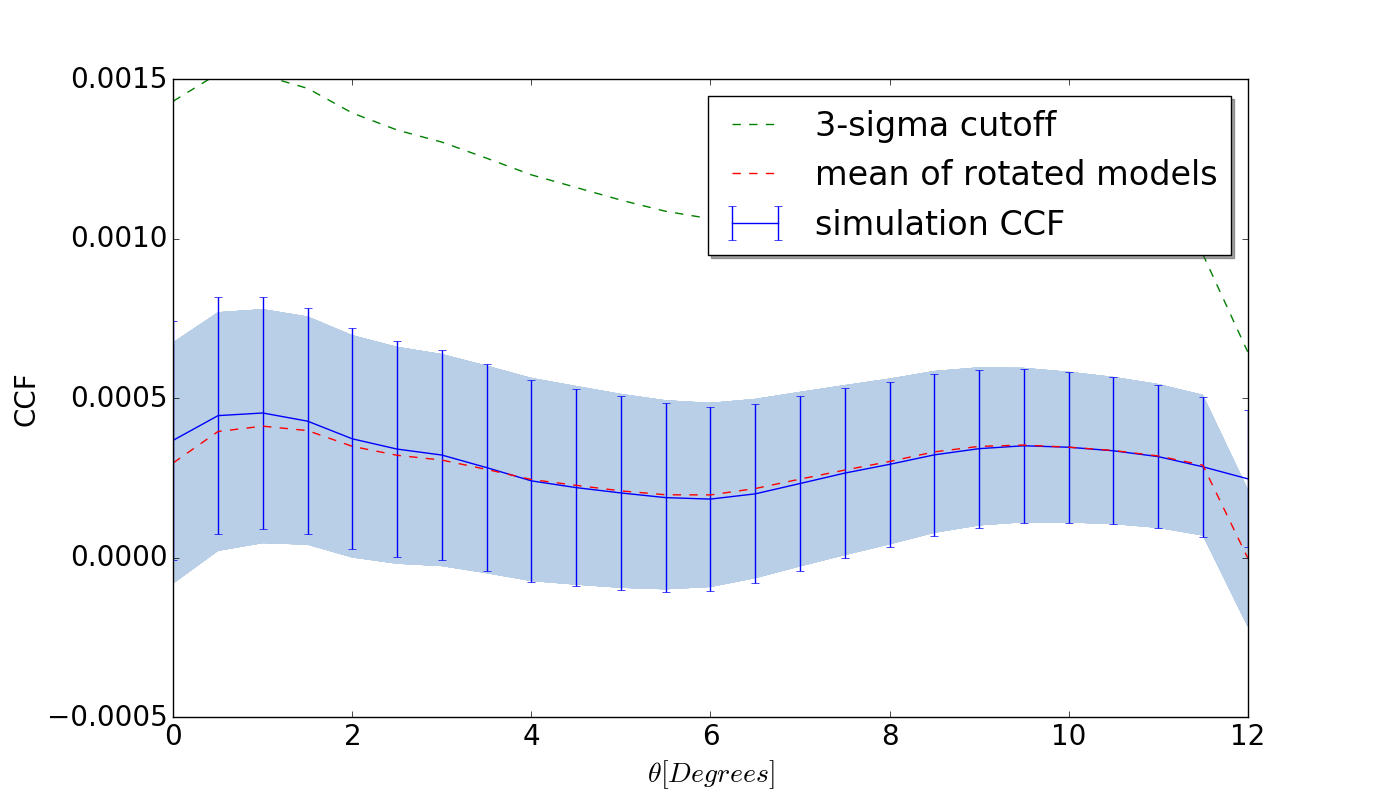}
\includegraphics[width=14cm]{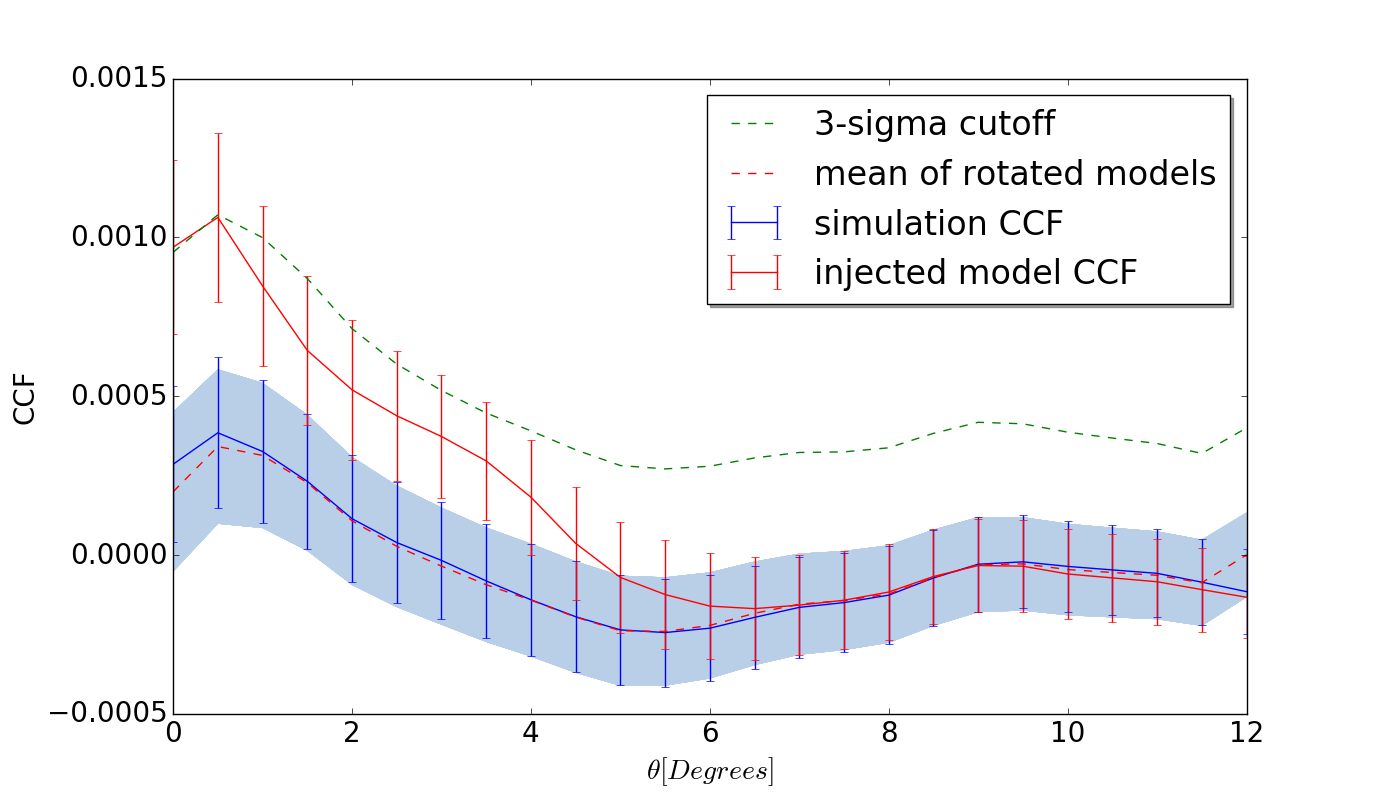}
\caption{\label{ccf} (top) The cross-correlation function for the {\it cosmological} model with the S-PASS radio map is the blue solid line. The blue shaded region is the $\pm$1$\sigma$ from the mean (red dashed line) of the rotated models correlations with the SPASS map. The green dashed line is the 3$\sigma$ upper threshold. Bottom: Correlation functions as above after injecting fake cosmic web emission into the S-PASS map. The solid red line is the cross-correlation function of the SPASS map and the rotated model that was injected. }
\end{center}   
\end{figure*}

In order to interpret the significance of the result we need to know the range of CCF values expected when there is no real correlation present. To do this we must cross correlate the S-PASS image with random maps (which should have no true correlation) that have the same physical characteristics as the LSS simulation, such as the same mean and clustering properties. In order to simulate random maps, while insuring that the statistics are as close as possible to the original, we have created 10 new versions of the {\it cosmological} model by rotating/flipping the all-sky map by combinations of 45$^{\circ}$, 90$^{\circ}$, and 180$^{\circ}$ in Galactic latitude and longitude. These rotated maps, which should no longer be correlated with the real radio sky, were also cross-correlated with the S-PASS image. Figure \ref{ccf} (top) shows the plot of the CCF as a function of angular scale. The blue line shows the CCF of S-PASS and the original {\it cosmological} model map.  The dashed red line is the mean CCF of S-PASS and the 10 rotated maps, the shaded blue region is the 1$\sigma$ range of the CCFs\footnote{The 1$\sigma$ from the mean is obtained by the quadrature sum of the rms of the 10 random maps and the cosmic variance in each angular bin.} and the dashed green line is the 3$\sigma$ upper envelope. Note that the correlation function is slightly above mean of the random maps at small angles, but well below even the 1$\sigma$ significance level, and is considered a non-detection. \\

\noindent {\sc Upper Limits:} To estimate an upper limit from this non-detection, we must add a fake cosmic-web signal into the S-PASS map (below the noise) and repeat the cross-correlation procedure. To do this, we take one of the rotated cosmic web maps and add it to the S-PASS map iteratively with an increasing multiplicative factor (we have chosen the map that has been rotated by 180$^{o}$ in both Galactic latitude and longitude). For each iteration we reproduced the cross-correlation, and terminated the procedure when the cross-correlation of the injected S-PASS map and the corresponding rotated cosmic web map reaches roughly 3$\sigma$ at any angular scale. The injection takes place before the spatial scale filtering, and we repeat the exact process of cross-correlating this injected map with the rotated model maps. The filtering will remove some of the injected signal, alter the S-PASS map, and thus affect the shapes of the correlation functions, all of which should be included when calculating the limit. We show the cross-correlation of the model-injected S-PASS map at the point of termination in Fig. \ref{ccf} (bottom), where the cross-correlation function with the injected model (times the scale factor) has reached 3$\sigma$ at an angular scale of $\theta \approx 1^{o}$ (solid red line in Fig. \ref{ccf}). We should note that the correlation persists at a low level out the 5 degrees of our 300$^{\prime}$ median-weight filter. From the radio map of the {\it cosmological} model (Fig. \ref{models} Right), we used the multiplicative factor at convergence (800$\pm$10) to create an upper limit map; had the radio emission from the cosmic-web been as bright as this map, we would have detected it at the 3$\sigma$ level. From this map we measured the average surface-brightness in filaments (defined as $3\times10^{-4} < n_e < 5\times10^{-3}$ [cm$^{-3}$]) to be 20.9~mJy~beam$^{-1}$, or 0.16~mJy~arcmin$^{-2}$. The density range corresponds to an overdensity range of $\delta$=3-50, typical of filaments and cluster outskirts \citep[e.g.,][]{vazz14}. Figure \ref{web} (top) shows the upper limit map, masking out all but the filamentary regions (in the $n_e$ range described above) where the average radio brightness was taken. 
 
\section{Discussion \& Conclusion} We now investigate what the limit for the {\it cosmological} model means for our knowledge of the magnetic fields in large-scale structure, as well as the prospects for detection of the cosmic web with upcoming surveys. 

\subsection{Magnetic Fields}  
\begin{figure}
\begin{center}  
\includegraphics[width=7.8cm]{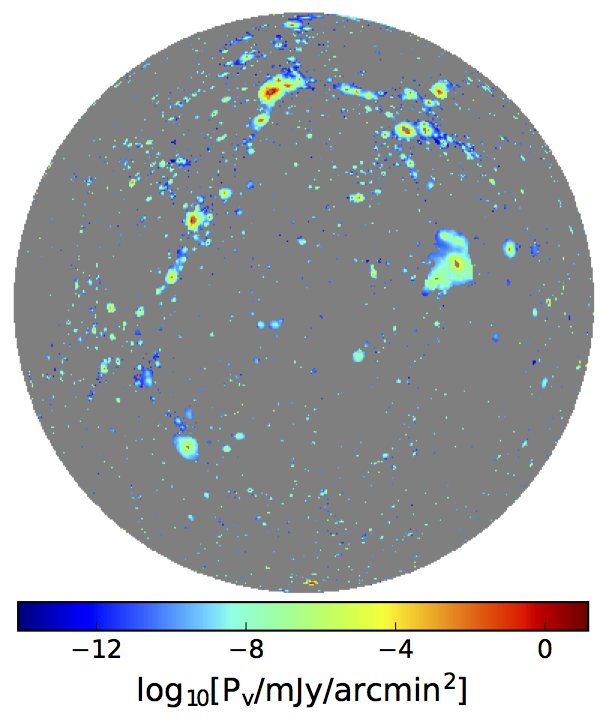}
\includegraphics[width=7.8cm]{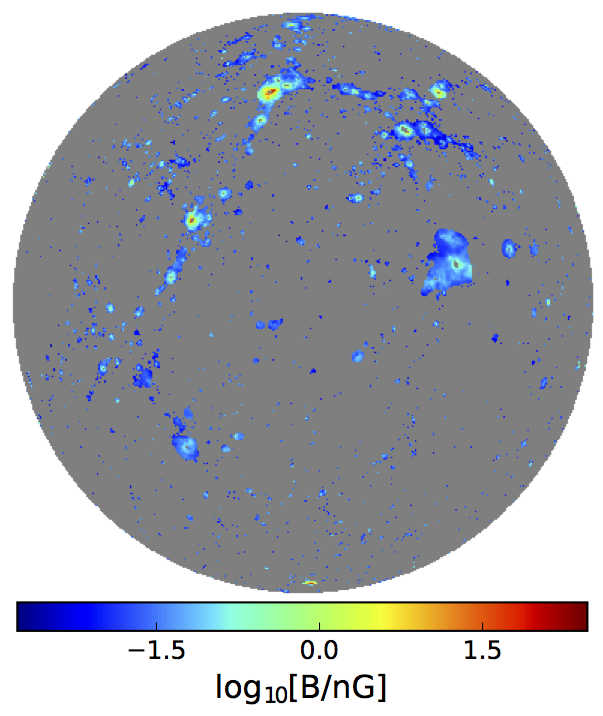}
\begin{footnotesize}
\caption{\label{web} Top: Orthographic projection of the upper limit {\it cosmological} model of the cosmic web with all structure except for the filaments masked out (see text). The mean synchrotron brightness in the filamentary structures is 0.16~mJy~arcmin$^{-2}$. Bottom: The magnetic field upper limit map derived from the same simulation, with the same mask. The average upper limit (mass-weighted) field is B$<$0.03 (0.13)~$\mu$G. }
\end{footnotesize}
\end{center}   
\end{figure}

To estimate the average magnetic field from the surface-brightness, we extracted the magnetic field values from the {\it cosmological} simulation (Fig. \ref{web}) and scaled it relative to the amount that the 3$\sigma$ upper limit surface-brightness is greater than the simulations brightness. In order to do this scaling, we must take into account both the spectral index of the radio emission $\alpha$ and the scaling of the synchrotron emissivity with the magnetic field. In the limit where the magnetic field B is much less than B$_{\mathrm{CMB}}$/(1+$z$)$^2$=3.24~$\mu$G (the equivalent effective magnetic field strength used to parameterize the electron Compton energy losses), the synchrotron emissivity $\epsilon_r$ of the simulation goes as 

\begin{equation}
\epsilon_r \sim X_{p} n_{e} B^{1+\alpha}~\nu^{-\alpha},
\end{equation} 
\noindent so we scale the magnetic field from the simulation by a factor of 

\begin{equation}
\left(\frac{(\epsilon_r^{\prime}/\epsilon_r)}{(\nu^{\prime}/\nu)^{-\alpha}}\right)^{1/(1+\alpha)},
\end{equation}

\noindent where $\epsilon^{\prime}$ is the average surface brightness of our injected signal,  $\nu^{\prime}$=2.3~GHz (the S-PASS frequency), $\nu$=1.4~GHz (the simulation frequency), $n_{e}$ is the number density of thermal electrons (in [cm$^{-3}$]), and $\alpha \approx$~1.0 for the secondary electrons assumed in the simulation. The factor $X_{p}$ is the relative fraction of the energy density of cosmic-ray protons with respect to the thermal gas, and for the current simulations a constant value of $X_{p}$=0.01 was used \citep{dola04}.  By taking the average value of the scaled magnetic field map in the same density regions as above, we obtain an upper limit magnetic field value of 0.03~$\mu$G. However, given that we are constraining the magnetic field using synchrotron emission, a density-weighted average of 
\begin{equation}
B_{\mathrm{filament}} < 0.13~\mu G~\left(\frac{0.01}{X_{p}}\right)^{1/(1+\alpha)} 
\end{equation}
\noindent over the region is more appropriate, where we have also included the dependency on $X_{p}$ and $\alpha$.  Fig. \ref{web} (bottom) shows a map of the magnetic field limits in the filamentary overdensity range where the average was taken. We should note that this value is dependent on the assumed model of the local magnetized cosmic web, especially the assumption that the CRe$^{\pm}$ were created through hadronic interactions, and does not include any primary cosmic-ray electrons produced in shocks and/or turbulence \citep[e.g.][]{vazz15a, vazz15b}. Not including these acceleration mechanisms makes this a true upper limit on the field, as simulations focusing only on primary electrons directly accelerated at structure formation shocks find comparable or higher densities of CRe$^{\pm}$ \citep[e.g.][]{skil08,aray12,vazz14,vazz15a}. Under the assumed model, we can use Equations (3) and (4) to infer a limit on the primordial magnetic field from the simulation to obtain 
\begin{equation}
B_{\mathrm{PMF}} < 1.0~nG ~\left(\frac{0.01}{X_{p}}\right)^{1/(1+\alpha)}. 
\end{equation}
\noindent To obtain this we take the assumed 10$^{-11}$ G PMF of the simulations, scale it by the factor in equation (3), keeping the model dependence explicit as in equation (4). This is a factor of a few below the limits obtained by Planck \citep[B$_{\mathrm{1 Mpc}} <$ 4.4 nG;][]{plan16}, and comparable to the recent limit derived through a joint Planck + South Pole Telescope analysis \citep[B$_{\mathrm{1 Mpc}} <$ 1.5 nG;][]{zucc16}.  However, constraints from the Coma cluster restrict $X_{p} \in [10^{-3}, 10^{-2}]$, which introduces a significant model uncertainty into Equ. (3) and (4) \citep[e.g.,][]{donn13}. 

\cite{vern17} recently performed a similar cross-correlation experiment using low frequency Murchison Widefield Array (MWA) data and the distribution of galaxy counts. The approach taken in \cite{vern17} uses deeper observations at lower frequency, both of which are advantages when searching for faint, steep spectrum emission \citep{brow11b}, though over a smaller area of the sky than presented here. Without access to a direct spatial model of the synchrotron cosmic web at higher redshift,  \cite{vern17} used a scaling relation derived from \cite{vazz15b} to find magnetic field limits of  0.03-1.98 $\mu$G, depending on the assumed spectral index, which is comparable to the limits derived here. While the \cite{vern17} work reached a deeper surface-brightness limit with the MWA, this work was able to achieve similar limits on the magnetic field by referencing a single simulated model of the local Universe. This in turn means that the limit presented here suffers from being highly dependent on the model of \cite{dola04}, in particular the source of the radiating CRe$^{\pm}$. Future observations that can independently constrain the cosmic-ray electron production mechanism can make observations of the cosmic web a competitive and complementary method for constraining primordial magnetic fields. 

\subsection{Future Direct Detection} The limits on the surface-brightness of the synchrotron cosmic web inferred above are deep compared to current radio observations due to the statistical reduction in the noise allowed by cross-correlation. The S-PASS limit derived at 2.3~GHz corresponds to $I_{\mathrm{1.4~GHz}}<$0.073~$\mu$Jy/arcsec$^{2}$, while the average brightness of diffuse synchrotron emission in clusters of galaxies is typically around 1~$\mu$Jy/arcsec$^{2}$ at 1.4~GHz \citep{brun14}. This brightness limit is not currently out of reach for deep, interferometric radio observations, though confusion from point-sources currently sets the limit on our ability to reach these levels in practice \citep{vern14,vern17}. \cite{vern17} demonstrated that confusion could be overcome statistically through cross-correlation, but direct detection will require detailed modeling and subtraction of detected compact sources. 

Upcoming all-sky surveys will reach sensitivities where direct detection of the brightest portions of the synchrotron cosmic web will be made possible. For example, the Evolutionary Map of the Universe \citep{norr11} survey with the Australian Square Kilometre Array Pathfinder (ASKAP), will have an RMS noise on the order of 0.1~$\mu$Jy/arcsec$^2$, and could potentially reach the above limit directly so long as the interferometric nature of ASKAP does not filter out the diffuse emission \citep{brow11b}. Given the optimistic value of X$_{p}$ assumed in this work, and the distance between our upper limit and the simulation magnetic field in Fig. \ref{mods}, the true radio emissivity in the cosmic-web might be out of reach for the EMU survey. See \cite{vazz15b} for a comparison of upcoming radio surveys (including LOFAR, MeerKAT, and the SKA) and their ability to probe the synchrotron cosmic web when primary accelerated electrons are considered. 

To summarize our results, we performed a cross-correlation between the S-PASS radio survey at 2.3~GHz with a simulated model of the local synchrotron cosmic web, and were able to place model dependent limits on the average surface-brightness ($<$0.16~mJy~arcmin$^{-2}$ at 2.3~GHz) and density-weighted magnetic field ($<$0.13~$\mu$G) at the 3$\sigma$ level. We further placed a limit on the primordial magnetic field (B$_{\mathrm{PMF}}<$1.0~nG) that is competitive to the current cosmic microwave background limits set by Planck + SPT \citep{zucc16}, highlighting the potential power of a direct detection of the synchrotron cosmic web with upcoming radio telescope surveys. 

\section*{Acknowledgements}
We would like to thank Franco Vazza for helpful discussions during the initial phases of this project. This work has been carried out in the framework of the S-band Polarisation All Sky Survey (S-PASS) collaboration. The Parkes Radio Telescope is part of the Australia Telescope National Facility, which is funded by the Commonwealth of Australia for operation as a National Facility managed by CSIRO. The Dunlap Institute is funded through an endowment established by the David Dunlap family and the University of Toronto. B.M.G. and T.V. acknowledge the support of the Natural Sciences and Engineering Research Council of Canada (NSERC) through grant RGPIN-2015-05948, and of the Canada Research Chairs program. Parts of this research were conducted" by the Australian Research Council Centre of Excellence for All-sky Astrophysics (CAASTRO), through project number CE110001020.

%\begin{figure*}
%\begin{center}  
%\includegraphics[width=16cm]{spass_cc_diagram.pdf}
%\begin{footnotesize}
%\caption{\label{flow} Diagram outlining the injection and cross-correlation procedure. }
%\end{footnotesize}
%\end{center}   
%\end{figure*}

%%%%%%%%%%%%%%%%%%%%%%%%%%%%%%%%%%%%%%%%%%%%%%%%%%



%\bibliographystyle{mnras}


%%%%%%%%%%%%%%%%%%%%%%%%%%%%%%%%%%%%%%%%%%%%%%%%%%

%%%%%%%%%%%%%%%%% APPENDICES %%%%%%%%%%%%%%%%%%%%%

%\appendix



%%%%%%%%%%%%%%%%%%%%%%%%%%%%%%%%%%%%%%%%%%%%%%%%%%

% Don't change these lines
\bsp	% typesetting comment
\label{lastpage}
\end{document}